\documentclass[twocolumn]{aastex631}
\usepackage{amsmath}
\usepackage{xcolor}

\usepackage{natbib}
\usepackage{booktabs}
\usepackage{hyperref}

\begin{document}

\title{Formation of a Possible Black-hole Ultracompact X-ray Binary with the Shortest Orbital Period}

\author[0009-0001-6764-106X]{Xing-Peng Yang}
\affil{School of Science, Qingdao University of Technology, Qingdao 266525, People’s Republic of China;chenwc@pku.edu.cn}
\author[0000-0002-9739-8929]{Kun Xu}
\affil{School of Science, Qingdao University of Technology, Qingdao 266525, People’s Republic of China;chenwc@pku.edu.cn}
\author[0000-0002-0138-3360]{Zhi-Fu Gao}
\affil{Xinjiang Astronomical Observatory, Chinese Academy of Sciences, Urumqi, Xinjiang 830011, People’s Republic of China}
\author[0000-0002-2479-1295]{Long Jiang}
\affil{School of Science, Qingdao University of Technology, Qingdao 266525, People’s Republic of China;chenwc@pku.edu.cn}
\affil{School of Physics and Electrical Information, Shangqiu Normal University, Shangqiu 476000, People’s Republic of China}
\author[0000-0002-0785-5349]{Wen-Cong Chen}
\affil{School of Science, Qingdao University of Technology, Qingdao 266525, People’s Republic of China;chenwc@pku.edu.cn}
\affil{School of Physics and Electrical Information, Shangqiu Normal University, Shangqiu 476000, People’s Republic of China}

\begin{abstract}
In the bulge of M31, the Chandra observations discovered a possible black hole (BH) ultracompact X-ray binary (UCXB) Seq.1 with an orbital period of $7.7$ minutes and a maximum X-ray luminosity $L_{\rm X}=1.09^{+0.02}_{-0.01}\times10^{38}~\rm erg\,s^{-1}$ in the $0.5-8$ keV band. The minimum orbital period of the BH UCXBs predicted by the standard magnetic braking (MB) model is longer than $8.3$ minutes. In this work, we investigate whether the convection- and rotation-boosted (CARB) MB prescription can account for the formation of a BH UCXB like Seq.1. Our detailed stellar evolution models indicate that the CARB MB law can drive isolated BH-main sequence (MS) binaries to evolve toward BH UCXBs with an orbital period of $7.7~\rm minutes$, in which a low-mass white dwarf transfers the material onto a BH in a short-term mass transfer episode, producing an X-ray luminosity of $\sim10^{38}~\rm erg\,s^{-1}$. We also obtain an initial parameter space of BH-MS binaries as the progenitors of Seq.1 in the donor-star masses and orbital periods plane, which can be applied to future population synthesis simulations. If Seq.1 is indeed a BH UCXB, future spaceborne gravitational wave (GW) detectors can detect the low-frequency GW signals from this source, and a tidal disruption event will be expected after $0.12$ Myr.

\end{abstract}

\section{Introduction}
Ultracompact X-ray binaries (UCXBs) are a subtype of low-mass X-ray binaries (LMXBs), which are characterized by an extremely short orbital period less than 1 hour \citep{taur23}. According to their compact orbits, the donor stars were inferred to be fully degenerate white dwarfs (WDs) or partially degenerate helium (He) stars \citep{rapp82,nels86,Pods02,delo03}. So far, there exists 20 identified UCXBs, which includes 11 persistent sources and 9 transient sources \citep{arma23}. Among those identified sources, the accreting compact objects in 19 sources are confirmed to be neutron stars (NSs). The unique candidate for accreting black holes (BH) is the luminous X-ray source X9 in globular cluster 47 Tucanae, in which a BH may be accreting from a WD in a close orbit \citep{mill15,bahr17,chur17,tudo18}. 

UCXBs are stellar and binary evolution fossils, which provide a lot of information in understanding the loss mechanisms of angular momentum \citep{ma09,haaf12a,haaf12b,chen16,chen21}, common envelope evolution \citep{zhu12}, mass transfer and accretion process \citep{lin18}, and the possible accretion-induced collapse of accreting NSs \citep{Chen23}. Furthermore, UCXBs are continuous low-frequency gravitational wave (GW) sources that could be detected by spaceborne GW detectors such as LISA, TianQin, and Taiji \citep{nele09,chen20a,chen20b,wang21,qin23,qin24}. Therefore, there exists a broad astrophysical significance for studying UCXBs.

Employing extensive Chandra observations in 2 million seconds spanning a temporal baseline of 16 yr, \cite{Zhang24} performed a systematic investigation for periodic X-ray sources in the M31 bulge. In their detected seven periodic X-ray sources, the source with sequence number 1 (Seq.1) has the highest X-ray luminosity (the unabsorbed luminosity is $L_{\rm X}=1.09^{+0.02}_{-0.01}\times10^{38}~\rm erg\,s^{-1}$ in the 0.5 to 8 keV band) and was naturally thought to be an active LMXB \citep{Zhang24}. According to an X-ray spectrum softer than that of NS X-ray binaries in the high hard state, the source was proposed to be a candidate BH X-ray binary \citep{Barn13}. The source Seq.1 was detected to have a period of 7.7 minutes \citep{Zhang24}. If it is indeed an orbital period, Seq.1 would be a candidate for UCXB with the shortest orbital period detected so far \citep{Zhang24}. Currently, the UCXB with the shortest orbital period is 4U 1728-34, which has an orbital period of $10.8~\rm minutes$ \citep{gall10}.

In dense clusters, UCXBs could be formed when BHs capture a companion star through dynamic processes including direct collisions, tidal captures, or exchange interactions \citep{Ivan05,Ivan10}. However, the dynamical processes are insignificant in a galactic field. To form UCXBs through the isolated binary evolution channel, a substantial loss of angular momentum is necessary during the evolution of their progenitor systems. The magnetic braking (MB) mechanism plays a vital role in driving the mass transfer of LMXBs with an orbital period longer than 3 hours \citep{Rapp83}. For relatively compact systems with an orbital period shorter than $2-3~\rm hours$, gravitational radiation (GR) is a dominant mechanism that extracts orbital angular momentum from orbital motion. As a consequence, the MB mechanism plays an important role in influencing the evolution of the UCXB progenitors. Adopting the standard MB prescription given by \cite{Rapp83}, \cite{qin23} explored the detectability of BH UCXBs evolving from BH-main sequence (MS) star channel as low-frequency GW sources and found that the maximum GW frequency is less than $0.004~\rm Hz$, which corresponds to a shortest orbital period longer than $8.3~\rm minutes$. Furthermore, the calculated mass transfer rate adopting the standard MB prescription is approximately an order of magnitude lower than the obseved one of the UCXB 4U 1916-053 \citep{Pods02}. Therefore, it is impossible to account for such an UCXB as Seq.1 for the standard MB prescription.

In principle, a star's rotation would influence the stellar winds' velocity. Furthermore, the surface magnetic field of a star scales with the dynamo number, and is related to the Rossby number, which depends on the star's angular velocity and the turnover time of convective eddies \citep{van19mn}. Therefore, the connection between a star's rotation and convection can drive a strong couple between the stellar wind and magnetic field, which then influences the MB process. Considering these effects, \cite{Van19} derived a convection and rotation-boosted (CARB) MB prescription, which can reproduce the observed mass-transfer rates, the orbital periods, the mass ratios, and donor stars' effective temperatures of those persistent LMXBs. Combined with binary population synthesis and a detailed binary evolution model, the CARB MB law was found to be successful in reproducing the observed parameters of all persistent NS LMXBs and binary pulsars \citep{deng21}.

Based on the CARB MB prescription, in this work we investigate whether a BH UCXB like Seq.1 with an orbital period of $7.7~\rm minutes$ could evolve from a BH-MS binary via the isolated binary evolution channel. In Section 2, we describe the stellar evolution code and the CARB MB law. Section 3 presents the results of our detailed stellar evolution model. Finally, we give a discussion and a brief summary in Sections 4 and 5, respectively.
\section{Binary evolution code and the CARB MB law}
\subsection{Binary Evolution Code}
To understand the formation and evolution of the BH UCXB candidate Seq.1, we utilize the binary module in the Modules for Experiments in Stellar Astrophysics \citep[MESA version r12115;][]{paxt11,paxt13,Paxton15,paxt18,paxt19} to simulate the evolution of BH-MS star binaries. The beginning of the evolution is a binary consisting of a BH (with a mass of $M_{\rm bh}$) and an MS star (with a mass of $M_{\rm d}$) in a circular orbit. The BH is thought to be a point mass, and the code only models the nuclear synthesis and evolution of the MS star with solar composition (i.e., $X = 0.7, Y = 0.28, Z = 0.02$). A detailed MESA model always proceeds for each binary system until the time step exceeds a minimum time step limit or the stellar age is older than the Hubble time (14 Gyr).

Once the MS star fills its Roche lobe, the surface H-rich material is transferred onto the BH at a rate $\dot{M}_{\rm tr}$. The accretion efficiency of the BH is assumed to be $f$, hence the accretion rate is $\dot{M}_{\rm acc}=f\dot{M}_{\rm tr}$. In this work, we take $f=1$. During the accretion, it is generally thought that the mass-growth rate of the BH is limited by the Eddington accretion rate $\dot{M}_{\rm Edd}$, i.e. $\dot{M}_{\rm bh}={\rm min}(\dot{M}_{\rm Edd},f\dot{M}_{\rm tr})$. The excess material in unit time ($\dot{M}_{\rm tr}-\dot{M}_{\rm bh}$) is assumed to be ejected from the vicinity of the BH, carrying away the specific orbital angular momentum of the BH. The Eddington accretion rate of a BH is given by
\begin{equation}
\dot{M}_{\rm Edd}=\frac{4\pi GM_{\rm bh}}{\kappa c\eta},    
\end{equation}
where $c$ is the speed of light in vacuo, $G$ is the the gravitational constant, $\kappa=0.2(1+X)$ ($X$ is the H abundance of the accreting material) is the Thompson-scattering opacity of electrons, and $\eta$ is the energy conversion efficiency of the accreting BH. When $M_{\rm bh}<\sqrt{6}M_{\rm bh,0}$, $\eta=1-\sqrt{1-(M_{\rm bh}/3M_{\rm bh,0})^2}$ \citep[where $M_{\rm bh,0}$ is the initial mass of the BH,][]{pods03}.

In the simulation, Type 2 opacities were used for extra C/O burning during and after He burning. Compared with Type 1 opacities with fixed metal distributions, Type 2 opacities support varying amounts of C / O beyond those accounted for by $Z$ \citep{igle96}. The ratio of the mixing length to the local pressure scale height is taken to be $\alpha=2$. Once the MS star fills its Roche lobe, the "Ritter" mass transfer scheme is arranged \citep{ritt88}. In the CARB MB law, the loss rate of stellar winds is an important input parameter (see also Section 2.2). we adopt the "Dutch" wind setting options with a scaling factor of 1.0 in the schemes including $hot_{-}wind_{-}scheme$ \citep{gleb09}. Our inlists are available on Zenodo:10.5281/zenodo.15289921.

During the evolution of BH binaries, we consider three loss mechanisms of orbital angular momentum. The total loss rate of orbital angular momentum is as follow
\begin{equation}
\dot{J}=\dot{J}_{\rm gr}+\dot{J}_{\rm ml}+\dot{J}_{\rm mb},   
\end{equation}
where $\dot{J}_{\rm gr}, \dot{J}_{\rm ml}$, and $\dot{J}_{\rm mb}$ are the loss rates of orbital angular momentum caused by GR, mass loss, and MB, respectively. It is emphasized that the stellar winds of the MS star play two roles: first, they carry away the specific orbital angular momentum of the donor star; second, they draw the spin angular momentum of the donor star via the MB process, resulting in a spin down of the star. However, the tidal interaction between two components would continuously spin the star back up to synchronous rotation with the orbit. The spin-up of the star consumes the orbital angular momentum, hence the MB mechanism indirectly extracts the orbital angular momentum.

\subsection{CARB MB prescription}

To produce a calculated mass-transfer rate that can match the observed one in LMXBs, \cite{Van19} proposed the CARB MB prescription. Assuming that the stellar winds are spherical symmetry, and are ejected at the $\text{Alfvén }$ radius ($R_{\rm A}$), the rate of angular momentum loss is 
\begin{equation}
  \dot{J}_{\rm mb} = -\frac{2}{3}\dot{M}_{\rm w}R_{\rm A}^{2}\Omega,  
\end{equation}
where $\dot{M}_{\rm w}$ and $\Omega$ are the wind-loss rate and the spin angular velocity (which is also the orbital angular velocity because an assumption of tidal lock) of the star, respectively. The $\text{Alfvén }$ radius $R_{\rm A}$ is tightly related to the surface escape velocity ($v_{\rm esc}$) of the star \citep{Matt12}. Considering the rotational effects of the star, the escape velocity can be modified to be
 \begin{equation}
     v'_{\rm esc}=(v_{\rm esc}^{2}+\frac{2\Omega^{2}R^{2}}{K^{2}})^{1/2},
 \end{equation}
where $R$ is the stellar radius, and $K=0.07$ is a constant fitted from a grid of simulations \citep{revi15}. 

Furthermore, the surface magnetic field of the star strongly depends on the rotation and the convective eddy turnover timescale \citep{park71,noye84,ivan06}. Including the influence of convection and rotation, the rate of angular momentum loss can be expressed as \citep{Van19}
\begin{align}
\dot{J}_{\rm mb}=-\frac{2}{3}\dot{M}_{\rm w}^{-1/3}R^{14/3}(v_{\rm esc}^2+2\Omega^2R^2/K^2)^{-2/3}\nonumber\\ \times\Omega_{\odot}B_{\odot}^{8/3}(\frac{\Omega}{\Omega_{\odot}})^{11/3}(\frac{\tau_{\rm conv}}{\tau_{\odot,\rm conv}})^{8/3},
\end{align}
where $\tau_{\rm conv}$ is the turnover time of convective eddies of the donor star, $\Omega_\odot \approx 3 \times 10^6 ~\rm s^{-1}$, $B_\odot = 1.0~ \rm G$, and $\tau_{\odot, \rm conv} = 2.8 \times 10^6 ~\rm s$ are the Sun's surface rotation rate, surface magnetic field strength, and convective turnover time, respectively. If the donor star has both a convective envelope and a radiative core, the CARB MB law is arranged to operate in the MESA code.

\begin{figure}
\centering
\includegraphics[width=1.0\linewidth,trim={0 0 0 0},clip]{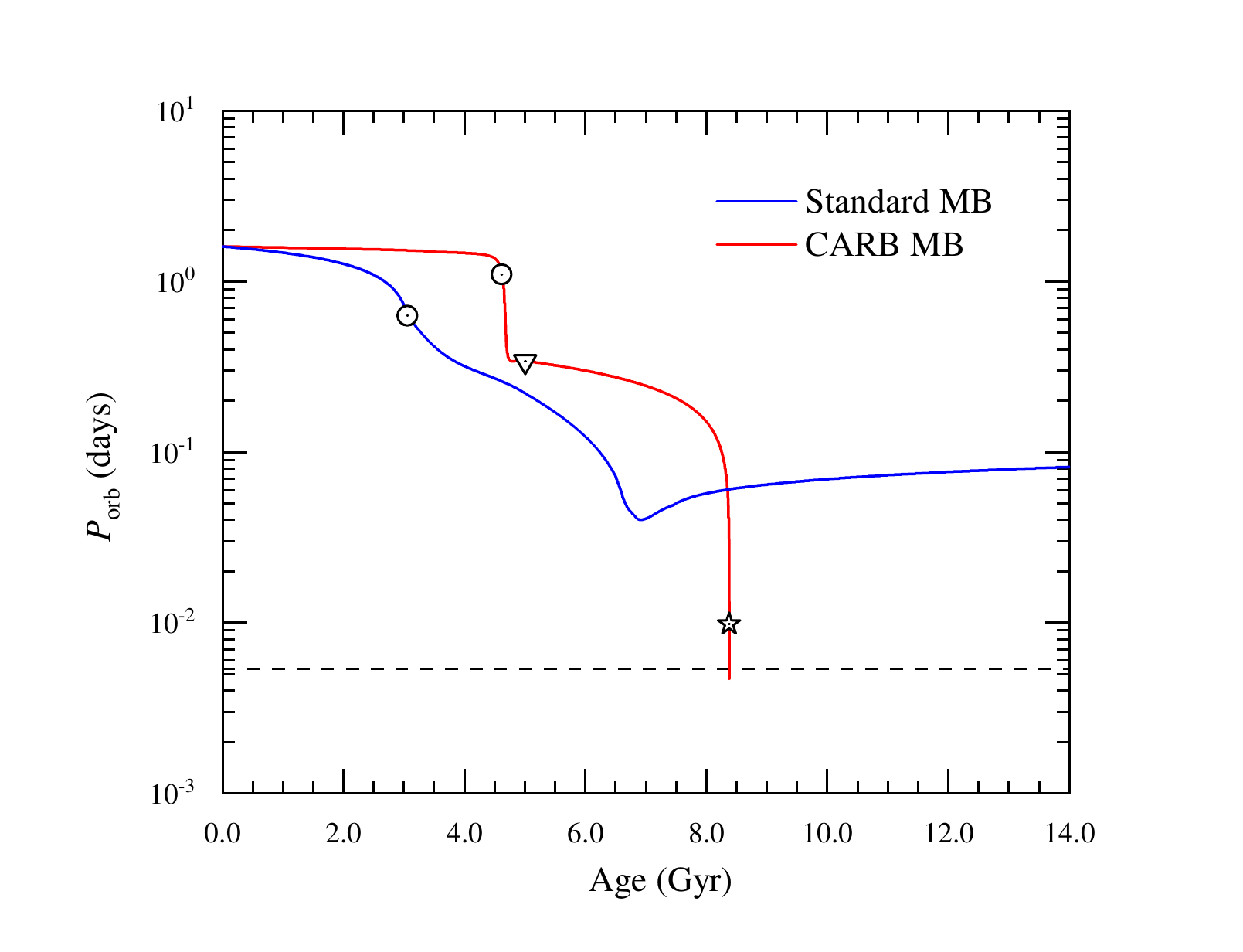}
\includegraphics[width=1.0\linewidth,trim={0 0 0 0},clip]{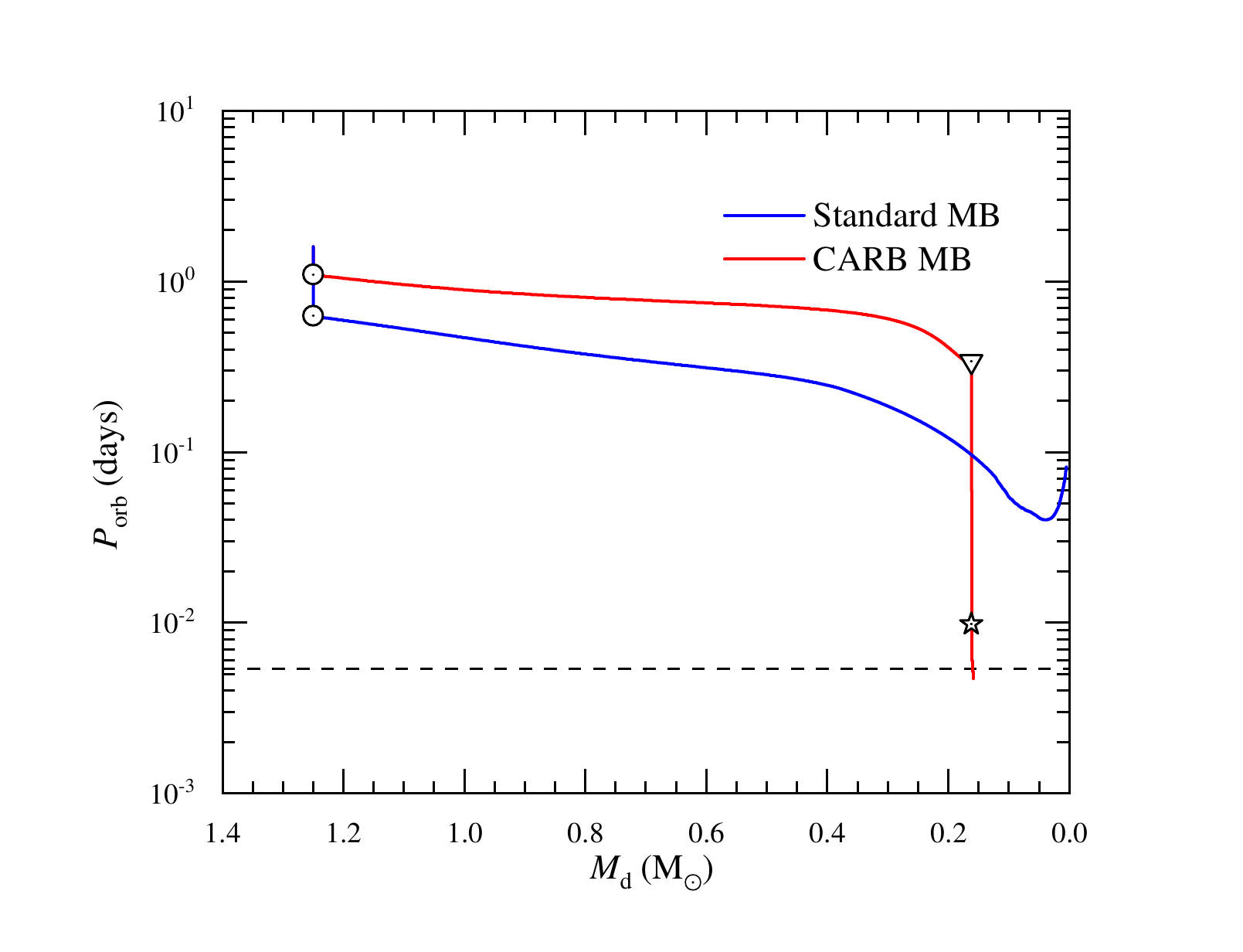}
\caption{Evolution of a BH-MS binary with an initial donor-star mass of  $1.25~M_{\odot}$, an initial BH mass of $10~M_{\odot}$, and an initial orbital period of $1.6~\rm days$ in the orbital period vs. stellar age diagram (top panel), and orbital period vs. donor-star mass diagram (bottom panel). The red and blue curves represent the simulated results given by the CARB and standard MB (with a MB index $\gamma=4$) prescriptions, respectively. The horizontal dashed lines indicate the observed period ($7.7~\rm minutes$) of Seq.1. The open circles, triangles, and stars denote the onset of the first mass transfer, the end of the first mass transfer, and the onset of the second mass transfer, respectively.
} \label{f1}
\end{figure}

\begin{figure}
\centering
\includegraphics[width=1.15\linewidth,trim={0 0 0 0},clip]{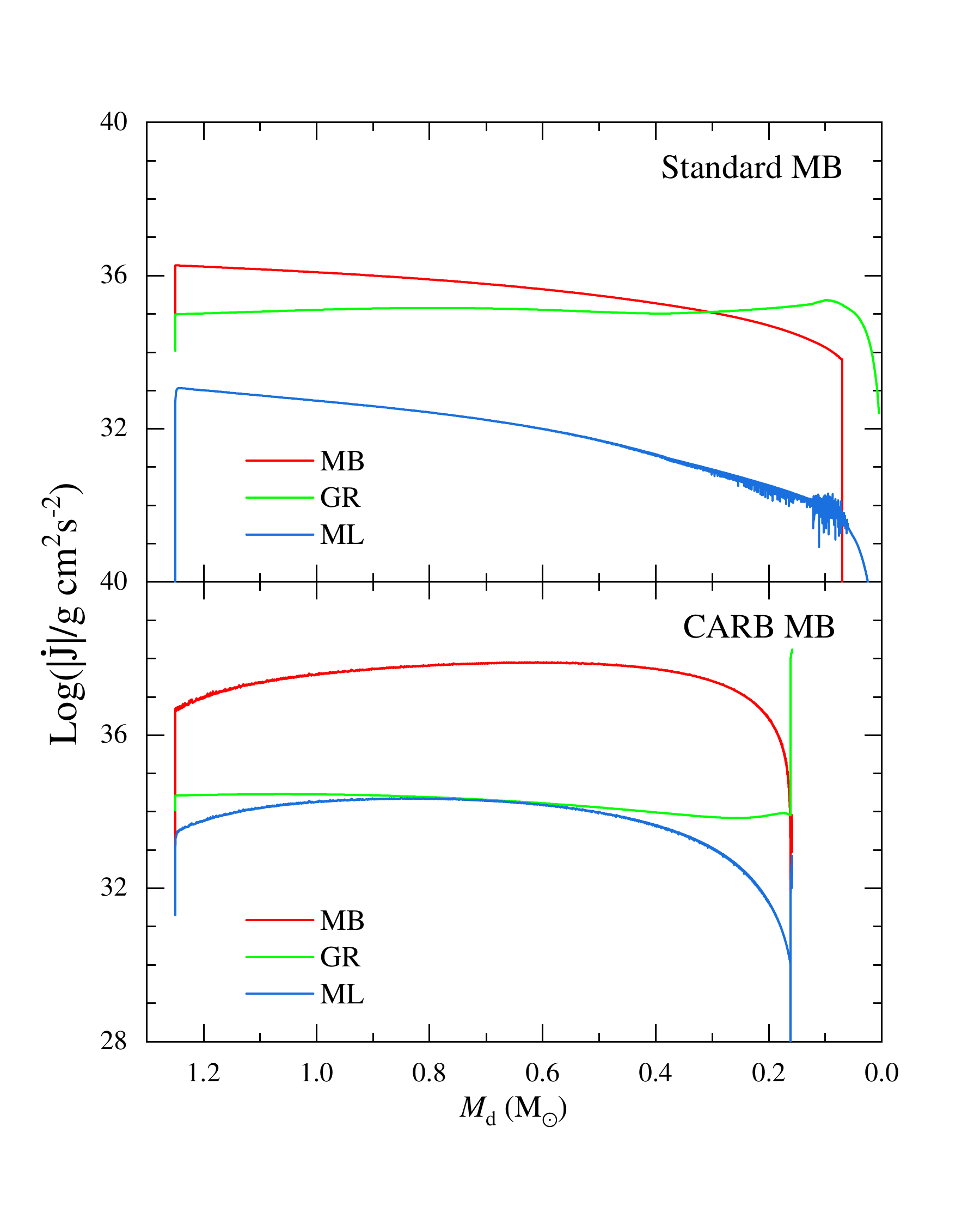}
\caption{Same as Figure \ref{f1}, but for loss rates of angular momentum by different mechanisms as a function of the donor-star masses. MB, GR, and mass loss are described by red, green, and blue curves, respectively. The top and bottom panels correspond to the standard and CARB MB cases, respectively.} \label{f2}
\end{figure}

\begin{figure}
\centering
\includegraphics[width=1.15\linewidth,trim={0 0 0 0},clip]{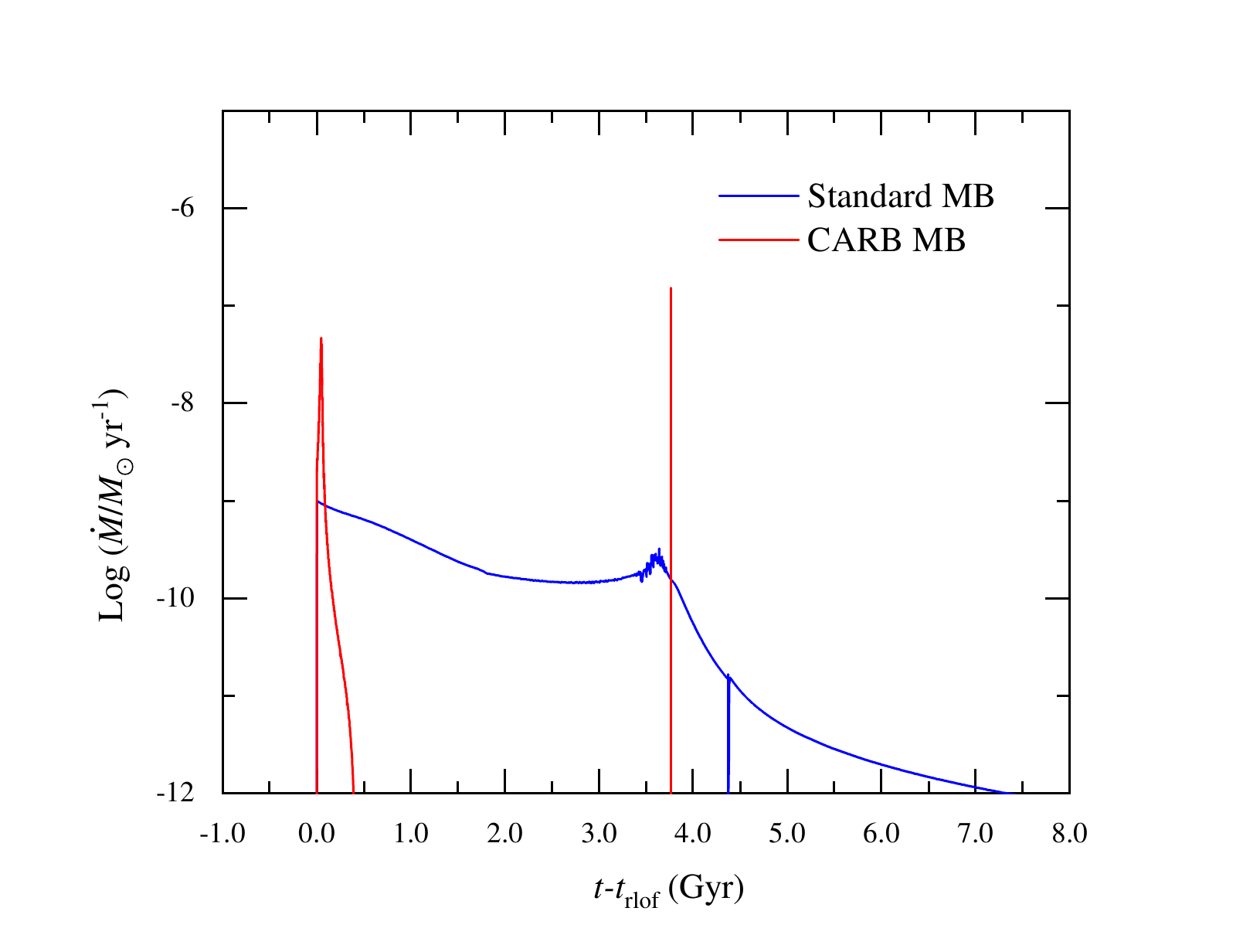}
\caption{Same as Figure \ref{f1}, but for the evolution of BH-MS binaries in the mass-transfer rate vs. mass-transfer timescale diagram. The red and blue curves correspond to the cases given by the CARB MB and the standard MB laws, respectively. In the figure, $t$ and $t_{\rm rlof}$ are the stellar age and stellar age when the RLOF occurs at
onset of LMXB, respectively. For this system, $t_{\rm rlof}=3.1~\rm Gyr$ and $4.6~\rm Gyr$ for the standard MB and CARB cases, respectively.
} \label{f3}
\end{figure}

\begin{figure}
\centering
\includegraphics[width=1.15\linewidth,trim={0 0 0 0},clip]{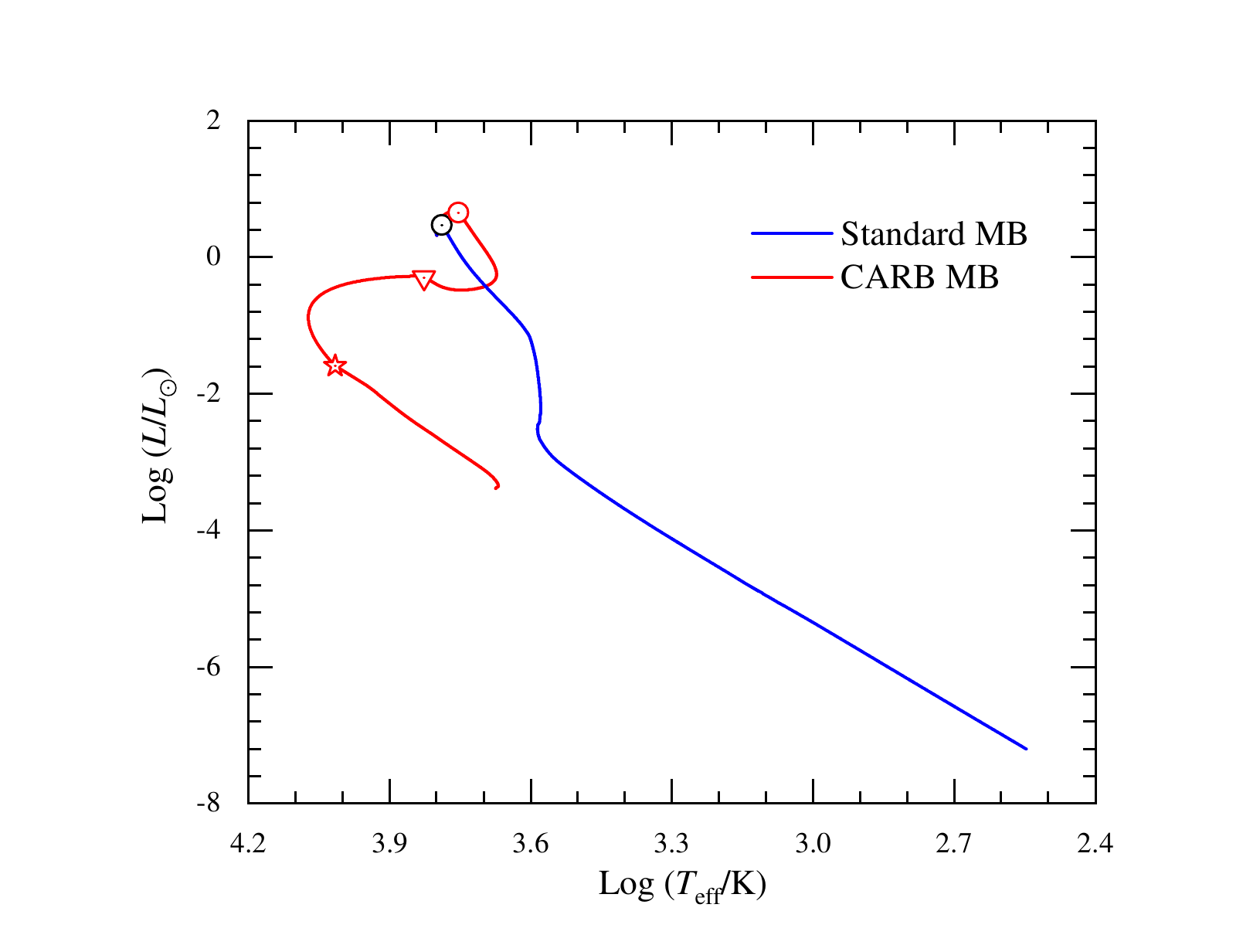}
\caption{Same as Figures \ref{f1} and \ref{f2}, but for the evolutionary tracks of BH-MS binaries in the H-R diagram.} \label{f4}
\end{figure}

\begin{figure}
\centering
\includegraphics[width=1.15\linewidth,trim={0 0 0 0},clip]{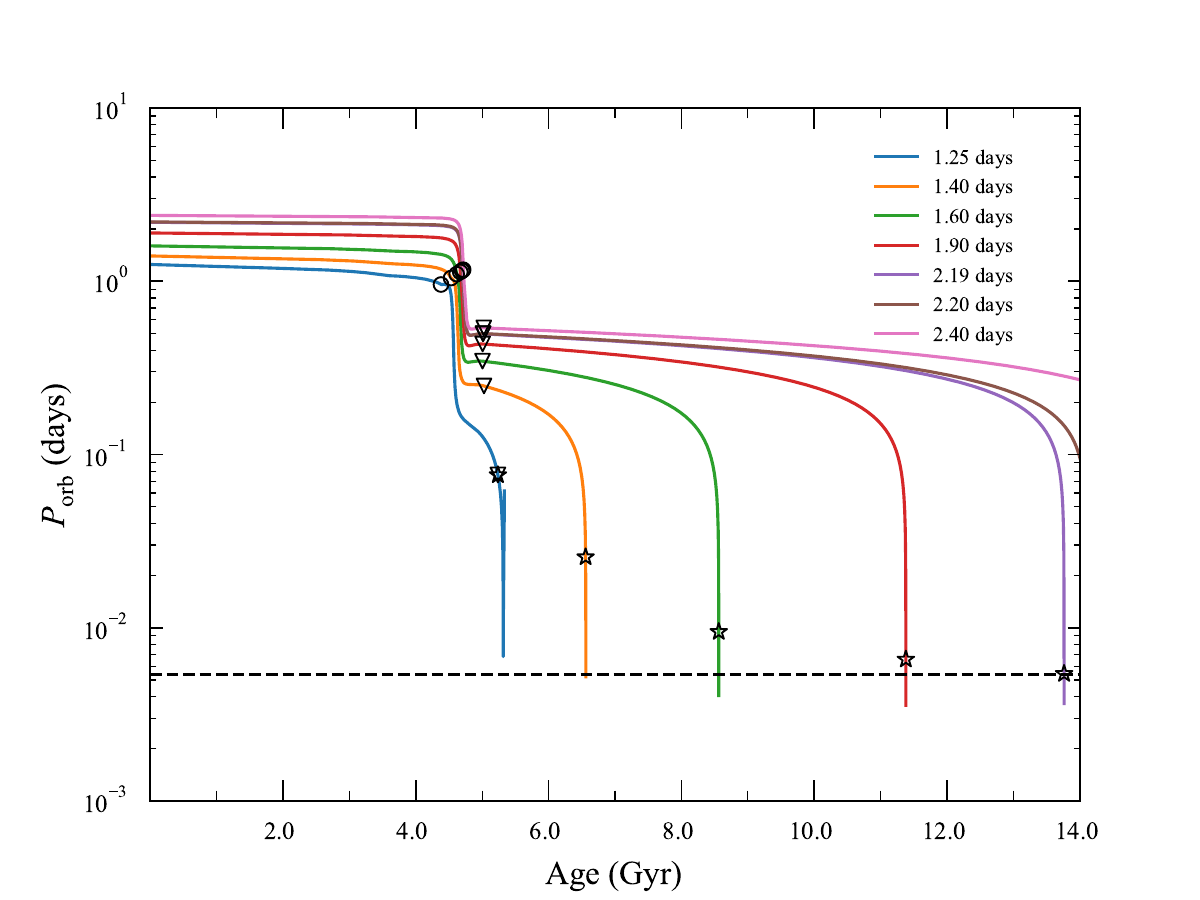}
\caption{Evolution of BH-MS binaries with a BH mass of $10~M_\odot$, a donor-star mass of $1.25~M_\odot$, and different initial orbital periods in the orbital period vs. stellar age diagram. The open circles, open triangles, and open stars represent the onset of first RLOF, the end of
first RLOF, and the onset of UCXBs, respectively. The horizontal dashed line denote the detected period ($7.7~\rm minutes$) of Seq.1.
} \label{f5}
\end{figure}

\section{Simulated Results}
\subsection{Some evolutionary examples}
Figure \ref{f1} plots the evolution of two BH-MS binaries with $M_{\rm bh}=10~M_{\odot}$, $M_{\rm d}=1.25~M_{\odot}$, and initial orbital period $P_{\rm orb,i}=1.60~\rm days$. At Roche lobe overflow (RLOF), the orbital periods are $0.62~\rm days$ and $1.09~\rm days$, and the abundances of central elements of the donor stars are $X_{\rm c}=0.19,Y_{\rm c}=0.79$ and $X_{\rm c}=0,Y_{\rm c}=0.98$ in the standard MB and CARB MB cases, respectively. These discrepancies arise from different loss rates of angular momentum before RLOF, in which the loss rate ($\dot{J}_{\rm mb}\sim4\times10^{34}~\rm g\,cm^2s^{-2}$) of angular momentum produced by the standard MB law is an order of magnitude higher than that ($\dot{J}_{\rm mb}\sim2\times10^{33}~\rm g\,cm^2s^{-2}$) given by the CARB MB law. When the MS star fills its Roche lobe, the surface H-rich material is transferred onto the BH through the inner Lagrange point. Since the mass is transferred from the less massive donor star to the more massive BH, the mass transfer would cause an orbital expansion effect. However, the orbital decay effect resulting from the angular momentum loss due to the MB process is stronger than the orbital expansion effect. As a consequence, the MB mechanism dominates the orbital evolution of the BH binary, resulting in a gradual orbital decay. In this stage, the evolutionary tracks of both the standard MB and the CARB cases exhibit a distinct difference. 

During the mass transfer, the evolution of loss rates of orbital angular momentum caused by GR, mass loss, and MB is designed in Figure \ref{f2}. After mass transfer begins, the CARB MB mechanism continuously becomes strong, and the loss rate of orbital angular momentum is $10^{37-38}~\rm g\,cm^2s^{-2}$. However, the loss rate of orbital angular momentum provided by the standard MB mechanism gradually decreases from $10^{36}$ to $\sim10^{35}~\rm g\,cm^2s^{-2}$. Similarly to \cite{bell23}, the MB process ceases when the donor star detaches from the Roche lobe and remains a He core in the CARB MB case. However, the MB mechanism in the standard MB case never stops until the donor-star mass decreases to $0.07~M_\odot$, at which the stellar core becomes completely convective.

In the standard MB case, the BH binary always appears as a BH X-ray binary until the donor-star mass decreases to $0.0054~M_{\odot}$ at $t=14~\rm Gyr$. \cite{qin23} already noticed that the BH-MS binaries cannot evolve into compact detached BH-WD systems when they adopt the standard MB law. Similarly, \cite{istr14} found that it is difficult to form detached neutron star-He WD systems in close orbits. Adopting the standard MB law with an unusual index of $\gamma=5$, it still required a severe fine-tuning for the initial orbital period of the progenitor systems \citep[see also Figures 2 and 6 of][]{istr14}. When $M_{\rm d}=0.31~M_\odot$ and $P_{\rm orb}=0.19~\rm days$, the GR begins to dominate the orbital evolution (see also Figure 2). Once the orbital expansion effect caused by the mass transfer exceeds the orbital decay effect caused by the GR, the orbital period reaches a minimum. In the standard MB case, the minimum orbital period is $58~\rm minutes$, which is much longer than the orbital period of $7.7~\rm minutes$ detected in Seq.1.

In the CARB MB case, the BH binary experiences rapid orbital decay and the orbital period decreases to $0.34~\rm days$ on a short timescale of $\sim 0.4~\rm Gyr$. As $M_{\rm d}=0.16~M_\odot$, the rich-H envelope of the donor star is almost stripped, remaining a He core, and the MB mechanism stops because of the absence of a radiative core. As a result, the donor star decouples from the Roche lobe, and the binary becomes a detached system. Subsequently, the low-mass He core would evolve into a He WD through a contraction and cooling phase. Meanwhile, the angular momentum loss caused by GR causes a continuous orbital decay until the He WD fills the Roche lobe again. Consequently, a short-term episode of mass transfer makes the BH binary appear as a UCXB with an orbital period of $7.7~\rm minutes$. Subsequently, the simulation stops because the time step exceeds a minimum time step limit. Before this, the time step declines rapidly due to the jump of the ratio ($v_{-}div_{-} csound_{-} surf$) between the rotation velocity and sound speed at outermost grid point of the donor star.

Figure \ref{f3} illustrates the evolution of the BH-MS binary of Figure \ref{f1} in the mass transfer rate versus mass transfer timescale diagram. At $t_{\rm rlof}=3.1$ and $4.6~\rm Gyr$, the donor stars fill the Roche lobes and initiate mass transfer at a rate of $\sim10^{-9}$ and $\sim10^{-8}~M_\odot\,\rm yr^{-1}$ for the standard MB and CARB MB cases, respectively. Before the RLOF, the loss rate of angular momentum given by the standard MB law is much higher than that in the CARB law; hence, the donor star in the former case fills the Roche lobe earlier than the latter. In the standard MB case, mass transfer proceeds at a rate of $\sim10^{-10}-10^{-9}~M_\odot\,\rm yr^{-1}$ on a timescale of $\sim3.9~\rm Gyr$. At a local minimum orbital period of $58~\rm minutes$, the mass-transfer rate reaches a peak. Subsequently, the mass transfer rate decreases sharply to $\sim10^{-12}-10^{-11}~M_\odot\,\rm yr^{-1}$. At $t-t_{\rm rlof}\approx4.4~\rm Gyr$, there exists a sudden drop in mass transfer, which is caused by a rapid shrinkage of the star that deviates from thermal equilibrium.

In the CARB MB case, the first mass transfer only lasts $\sim 0.4~\rm Gyr$ due to a relatively high mass transfer rate ($\sim10^{-8}~M_\odot\,\rm yr^{-1}$). With exhaustion of the H-rich envelope of the donor star, the remaining He core begins to evolve into a WD after a long-term ($\sim 3.3~\rm Gyr$) contraction and cooling phase. During the formation of the He WD, the continuous GR drives the BH-WD binary to evolve toward a compact orbit. Once the WD fills its Roche lobe, it triggers a short-term episode of mass transfer, in which a mass transfer rate of $\ge 10^{-9}~M_\odot\,\rm yr^{-1}$ ($\ge 10^{-8}~M_\odot\,\rm yr^{-1}$) lasts for a timescale of $0.16~\rm Myr$ ($0.07~\rm Myr$) before the simulation stops. If the mass transfer rate is $1.7\times 10^{-8}~M_\odot\,\rm yr^{-1}$, the X-ray luminosity of the accreting BH can be estimated to be $L_{\rm X}=0.1\dot{M}c^2\approx1.0\times10^{38}~\rm erg\,s^{-1}$, which is consistent with the highest X-ray luminosity observed in Seq.1. 

To understand the evolutionary properties of the donor stars,
Figure \ref{f4} displays their evolutionary tracks in the H-R diagram. After ${\rm log}(L/L_\odot)\sim-0.4$, the evolution of BH-MS binaries in two MB cases exhibits different tendency, in which the donor stars in the standard MB and the CARB MB cases evolve toward a low effective temperature and high effective temperature, respectively. This difference originates from various orbital evolutions in which the CARB MB law drives the BH-MS binary to evolve toward a close orbit after the first mass transfer begins. Because donor stars always fill the Roche lobe, its radius sharply decreases in the case of the CARB MB. According to the Stefan-Boltzmann law $L=4\pi R^2 \sigma T_{\rm eff}^4$ ($\sigma$ and $T_{\rm eff}$ are Stefan-Boltzmann constant and stellar effective temperature), a similar luminosity (${\rm log}(L/L_\odot)\sim-0.4$) naturally produces an increasing effective temperature. In the standard MB case, the donor-star radius slowly decreases, hence a decreasing luminosity causes the effective temperature to decline. In the CARB MB case, the donor star fills its Roche lobe again at ${\rm log}(L/L_\odot)\sim -2$ and $T_{\rm eff}\sim 10000~\rm K$, which are in the typical range of a WD. Subsequently, the WD begins a cooling phase like in the standard MB case.

Apart from the loss mechanisms of angular momentum, initial orbital periods also play a vital role in determining whether those BH binaries can evolve toward UCXBs. Adopting the CARB MB prescription, Figure \ref{f5} depicts the evolution of BH-MS binaries with $M_{\rm d}=1.25~M_{\odot}$, $M_{\rm bh}=10~M_{\odot}$, and different initial orbital periods in the orbital period versus stellar age diagram. Our simulations indicate that those BH-MS binaries have a bifurcation period of 2.19 days, which is defined as the longest initial period that pre-LMXBs can evolve into UCXBs within the Hubble timescale \citep{sluy05a,sluy05b}. For an initial orbital period of 2.2 days, the MB and GR can drive the binary to evolve toward a system with a compact orbit, but the evolutionary timescale exceeds the Hubble timescale.  When the initial periods are in the range from 1.40 to 2.19 days, those BH-MS star binaries first evolve into detached BH-WD systems. With orbital decay caused by the GR, the WDs then fill the Roche lobes, and the systems evolve into UCXBs with an ultra-short period of $7.7~\rm minutes$. However, it is impossible to evolve to the present period of Seq.1 as the initial period is 1.25 days. The reason caused this phenomenon is as follow: a shorter initial orbital period corresponds to a shorter evolutionary timescale before the RLOF, naturally developing a lower-mass He core in the stellar core; once the rich-H envelope of the donor star is stripped, the lower-mass He core evolves into a He WD, and the binary becomes a detached system consisting of a BH and a WD; meanwhile, the lower-mass He WD naturally has a larger radius, resulting in a longer orbital period when the He WD fills the Roche lobe. Therefore, it requires a specific initial orbital period range for the BH binaries to evolve toward Seq.1, in which the longest initial period is the bifurcation period, and the shortest initial period should be longer than a critical period.

\begin{figure}
\centering
\includegraphics[width=1.15\linewidth,trim={0 0 0 0},clip]{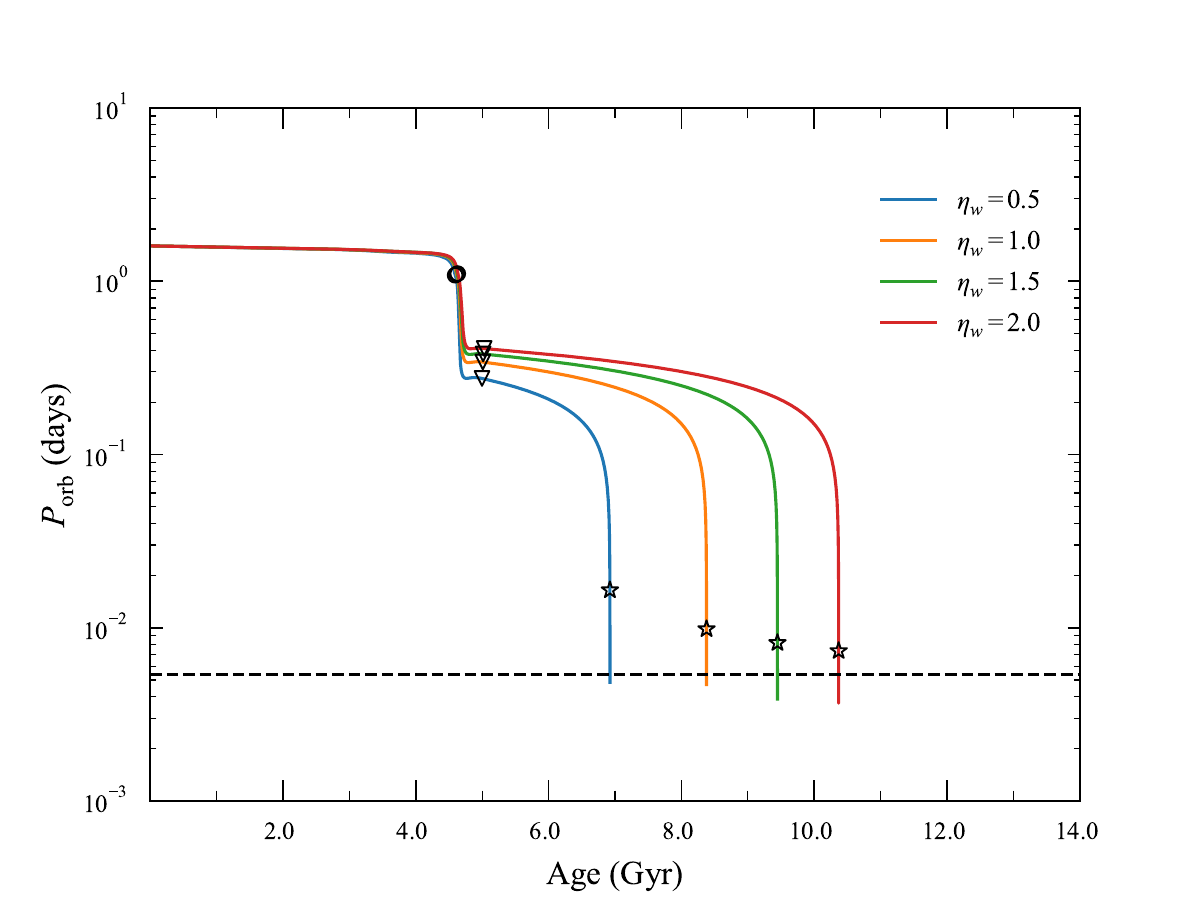}
\caption{Evolution of a BH-MS binary with $M_{\rm bh,i}=10~M_\odot$, $M_{\rm d,i}=1.25 ~M_\odot$, and $P_{\rm orb,i}=1.6 ~\rm days$ under different wind scaling factors in orbital period vs. stellar age diagram. The blue, yellow, green, and red curves correspond to a wind scaling factor of $0.5$, $1.0$, $1.5$, and $2.0$, respectively. The horizontal dashed line indicates the observed period (7.7 minutes) of Seq.1.} \label{f6}
\end{figure}

The wind scaling factor ($\eta_{\rm w}$) can determine the wind loss rate and influence the rate of angular momentum loss by the CARB MB mechanism. Figure \ref{f6} gives the evolution of BH-MS binaries with $M_{\rm bh,i}=10~M_\odot$, $M_{\rm d,i}=1.25~M_\odot$, and $P_{\rm orb,i}=1.6~\rm days$ under different $\eta_{\rm w}$ in $M_{\rm d}-P_{\rm orb}$ diagram. Four BH-MS binaries with a wind scaling factor of $0.5-2.0$ can evolve into Seq.1-like systems. When the BH-MS binaries evolve into detached systems, the system with a smaller wind scaling factor has a shorter orbital period. This is because a smaller wind scaling factor produces a lower wind loss rate, naturally resulting in a stronger MB mechanism according to equation (5).  Subsequently, the WD in the system with a smaller wind scaling factor fills the Roche lobe in a longer orbital period.

\begin{table*} [htbp]  
\caption{Some Important Evolutionary Parameters for those BH-MS Binaries with a BH of $10~M_\odot$ and a bifurcation period}
\begin{tabular}{ccccccccccc}  
\toprule  
$M_{\rm d,i}$ & $P_{\rm orb,i}$ & $t_{\rm rlof}$    & $t_{\rm deta}$ &  $P_{\rm orb,1}$ & $M_{\rm d,1}$ & $t_{\rm ucxb}$ &  $P_{\rm orb,2}$& $P_{\rm orb,min}$&$\Delta t_{-9}$& $\Delta t_{-8}$\\ 
$(M_{\odot})$ & $(\rm days)$ & $(\rm Gyr)$    & $(\rm Gyr)$ & $(\rm days) $ & $(M_{\odot})$ & $(\rm Gyr)$ & $(\rm min)$& $(\rm min)$&$(\rm Myr)$& $(\rm Myr)$\\  
\midrule
1.00 & 4.10 &11.458 &11.908 &0.271   &0.157  &13.843    &28.1541  &7.1063   &1.407  &0.155 \\
1.10 & 3.60 & 7.876 &8.233 &0.409   &0.165  &13.765    &10.1889  &5.9206   &0.060  &0.035 \\
1.20 & 2.45 &5.524 &5.867 &0.423   &0.166  &11.829    &9.7763  &5.5735   &0.056  &0.035 \\
1.30 & 1.40 &3.945 &4.334 &0.397   &0.165  &9.354    &10.7970  &5.9901   & 0.071  &0.040 \\
1.40 & 1.15 &2.707 &3.348 &0.398   &0.166  &8.386    &11.0323  &5.4513   &0.084  &0.050 \\
1.50 & 1.05 &1.960 &2.662 &0.436   &0.168  &8.794    &10.1126  &5.4696   &0.059  & 0.036 \\
1.60 & 0.75 &0.745 &2.599 &0.327   &0.161  &5.559    &13.5986  &6.6552   &0.148  &0.064 \\
1.70 & 0.65 &0.336 &2.597 &0.278   &0.157  &4.545    &20.0186  &6.8719   &0.576  &0.131 \\
1.80 & 0.60 &0.151 &2.404 &0.389   &0.164  &6.970    &10.0913  &5.7454   &0.060  &0.038 \\
\bottomrule  
\end{tabular} 
 \label{tab1}
\tablenotetext{}{Notes. 1. The columns (from left to right): the initial donor-star mass, initial orbital period, stellar age at the onset of RLOF, stellar age when the first mass transfer ceases, orbital period and donor-star mass when the first mass transfer ceases, stellar age at the onset of UCXB stage, orbital period at the onset of UCXB stage, minimum orbital period, timescales for the second mass transfer at a rate exceeding $10^{-9}$ and $10^{-8}~M_\odot\,\rm yr^{-1}$, respectively.\\
2. The simulation of all UCXBs stops because the time step reaches a minimum time step limit.}
\end{table*}

\subsection{Parameter space of BH-MS binaries forming Seq.1}
\begin{figure}
\centering
\includegraphics[width=1.15\linewidth,trim={0 0 0 0},clip]{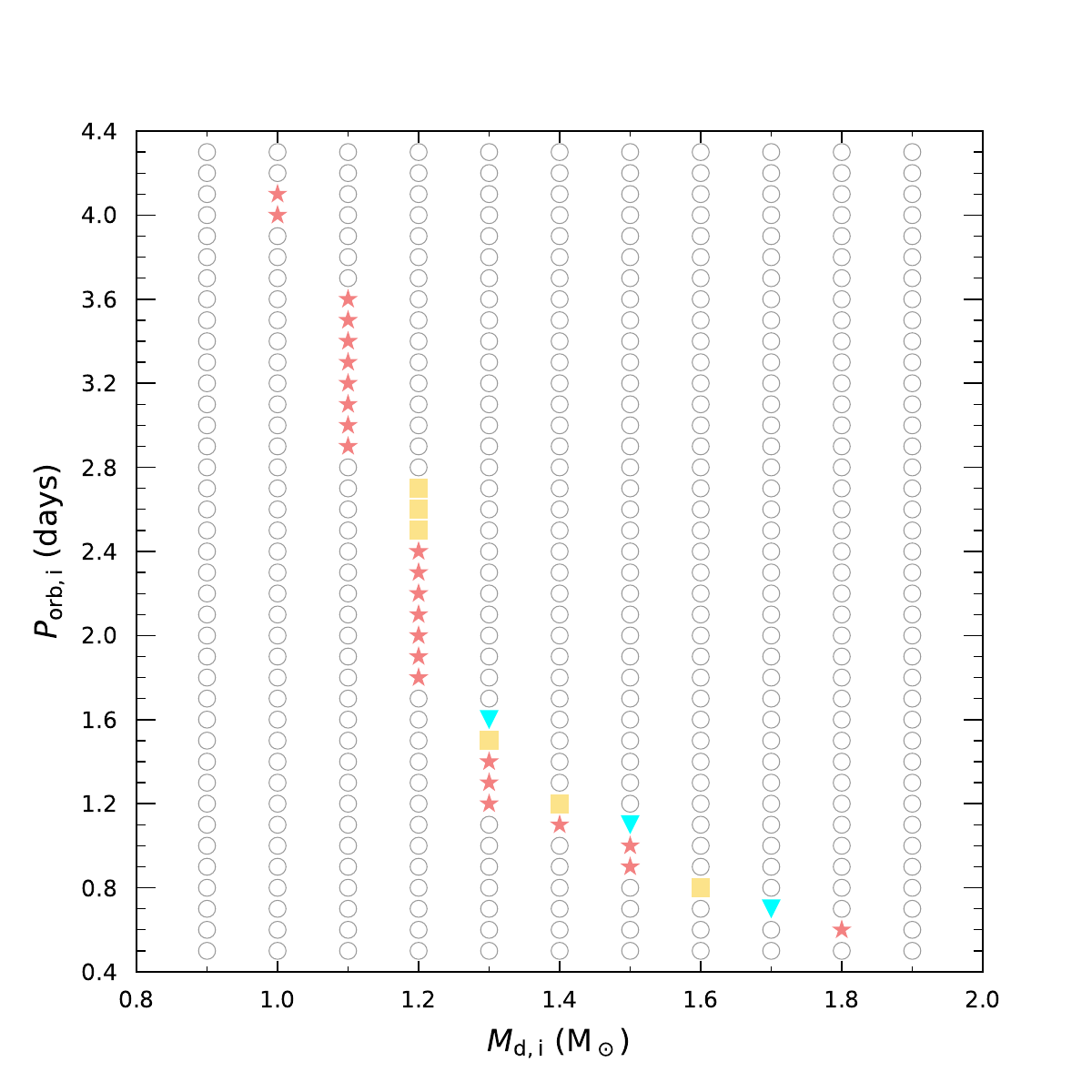}
\caption{Parameter space distribution of BH-MS binaries that can evolve toward 
BH UCXBs like Seq.1 in the initial orbital period vs. initial donor-star mass diagram. The initial mass of the BH is fixed to be $10~M_{\odot}$. The open circles indicate those systems that cannot evolve into UCXBs within a Hubble timescale, while the stars represent those BH-MS binaries that can evolve into BH UCXBs with a minimum orbital period less than $7.7~\rm minutes$. The current mass transfer rates in the red stars, yellow squares, and blue triangles are $>10^{-8}~M_{\odot}\, \rm yr^{-1}$,  $10^{-9}~M_{\odot}\, {\rm yr}^{-1}<\dot{M}_{\rm tr}<10^{-8}~M_{\odot}\, \rm yr^{-1}$, and $< 10^{-9}~M_{\odot}\, \rm yr^{-1}$, respectively.
} \label{f7}
\end{figure}

To understand the properties of the progenitors of Seq.1, we have simulated the evolution of a great number of BH-MS binaries. Figure \ref{f7} summarizes the initial parameter space of BH-MS binaries that can evolve toward Seq.1 in the $M_{\rm d,i}-P_{\rm orb,i}$ plane. Those BH-MS binaries with initial donor-star masses of $1.0-1.8~M_\odot$ and initial orbital periods of $0.6-4.1~\rm days$ could potentially evolve toward BH UCXBs like Seq.1. The upper boundary originates from the bifurcation periods, over which those BH-MS binaries would evolve toward the systems with wide orbits. Certainly, the bifurcation periods are tightly related to the initial donor-star masses. The BH-MS binaries with initial orbital periods shorter than the bottom boundary can evolve into compact orbit systems, but their orbital periods cannot reach $7.7~\rm minutes$. Since a lower-mass donor star requires a longer evolutionary timescale, those BH-MS binaries with a donor star lower than $1.0~M_\odot$ cannot evolve toward Seq.1 in a Hubble timescale. Those BH-MS binaries with initial donor-star masses higher than the right boundary can not evolve toward UCXBs because of the absence of MB process. 

It is worth emphasizing that those BH-MS binaries with a $1.6-1.8~M_\odot$ donor star can also evolve toward Seq.1. In the traditional MB scheme, those donor stars with a mass greater than $1.5~M_\odot$ are not expected to experience an MB process due to the lack of a convective envelope \citep{verb81,Rapp83}, hence the systems first evolve toward a relatively wide orbit. If the initial periods are not too long, the MB mechanism can still drive the binaries to evolve into short-period systems once the donor stars develop a convective envelope. Adopting the standard MB laws, \cite{qin23} obtained the initial parameter space of BH-MS binaries that can evolve toward UCXBs, in which the highest initial donor-star mass is $1.6~M_\odot$. In the CARB MB case, the highest initial donor-star mass is $1.8~M_\odot$, implying that the loss rate of angular momentum provided by the CARB MB law is much higher than that given by the standard MB law.

In principle, the initial BH mass would influence the parameter space of the progenitors of Seq.1. For the same donor star, a higher BH mass in those BH-MS binaries tends to evolve toward detectable LISA sources in a narrow initial orbital period range \citep{qin23}. A higher BH mass results in a lower mass ratio and a stronger orbital expansion effect due to a mass transfer. Furthermore, a higher BH mass produces a larger orbital angular momentum and a weaker orbital shrinkage effect under the same loss rate of angular momentum due to an MB process. As a result of competition between orbital expansion and orbital shrinkage effect, a higher BH mass naturally produces a system with a wide orbit \citep[for a details see also][]{qin23}. Therefore, a high BH mass tends to reduce the initial parameter space of the progenitors of Seq.1. Our detailed stellar evolution model found that both the upper and bottom boundaries of the parameter space move up in Figure \ref{f7} when $M_{\rm bh,i}=6~M_\odot$ (however, the total parameter space is slightly larger than that in $M_{\rm bh,i}=10~M_\odot$).

Table 1 lists some important evolutionary parameters of BH-MS binaries that can evolve toward BH UCXBs like Seq.1. The initial orbital periods of all BH-MS systems are very close to the bifurcation periods. It is clear that a lower initial donor-star mass corresponds to a longer bifurcation period. The underlying reason is as follows: a lower-mass donor star has a deeper convective envelope on the MS stage, resulting in a more efficient MB mechanism; meanwhile, a longer nuclear evolution timescale on the MS stage develops a more massive He core, which is beneficial to evolve toward a more compact orbit.  When the first mass transfer ends, the orbital periods are distributed in a narrow range of $0.271-0.436~\rm days$. Meanwhile, the WD masses in those detached BH-WD binaries concentrate within a narrow range of $0.157–0.168~ M_\odot$ which is very similar to the WD mass distribution in those detached NS-WD systems \citep{taur18,chen20b}. The minimum orbital periods in all systems are shorter than the detected period ($7.7~\rm minutes$) in Seq.1, and the shortest period is $5.5~\rm minutes$. During the second mass transfer, the longest and shortest timescales are $1.41$ ($0.16$) and $0.06~\rm Myr$ ($0.04~\rm Myr$) when the mass transfer rates exceed $10^{-9}~M_\odot\,\rm yr^{-1}$ ($10^{-8}~M_\odot\,\rm yr^{-1}$), respectively.

\section{Discussion}

\subsection{Detectability of Seq.1 as a continuous low-frequency GW sources}

If Seq.1 is indeed a BH UCXB, it is possible to detect its low-frequency GW signals for those future space-borne GW detectors. Adopting a mission duration of 4 yr, the GW characteristic strain of a BH-WD binary can be written as \citep{chen20a}
\begin{equation}
h_{c}\approx3.75\times10^{-20}\left(\frac{f_{\rm gw}}{0.001~\rm Hz}\right)^{7/6} \left(\frac{\mathcal{M}}{1~M_\odot}\right)^{5/3} \left(\frac{10~\rm kpc}{d}\right).
\end{equation}
where $f_{\rm gw}=2/P_{\rm orb}$ is the GW frequency and $d$ is the distance of the GW source. The chirp mass $\mathcal{M}$ is given by
\begin{equation}
\mathcal{M}=\cfrac{(M_{\rm bh}M_{\rm d})^{3/5}}{(M_{\rm bh}+M_{\rm d})^{1/5}}.
\end{equation}
We take the distance of 785 kpc for M31 as an estimated distance of Seq.1 \citep{mcco05}. In numerical calculations, if the simulated characteristic strain exceeds the sensitivity curve given by \cite{Robson19} for LISA, the corresponding binary system is considered a LISA source. For TianQin and Taiji two low-frequency GW detectors, we also take the same way like LISA.

\begin{figure}
\centering
\includegraphics[width=1.15\linewidth,trim={0 0 0 0},clip]{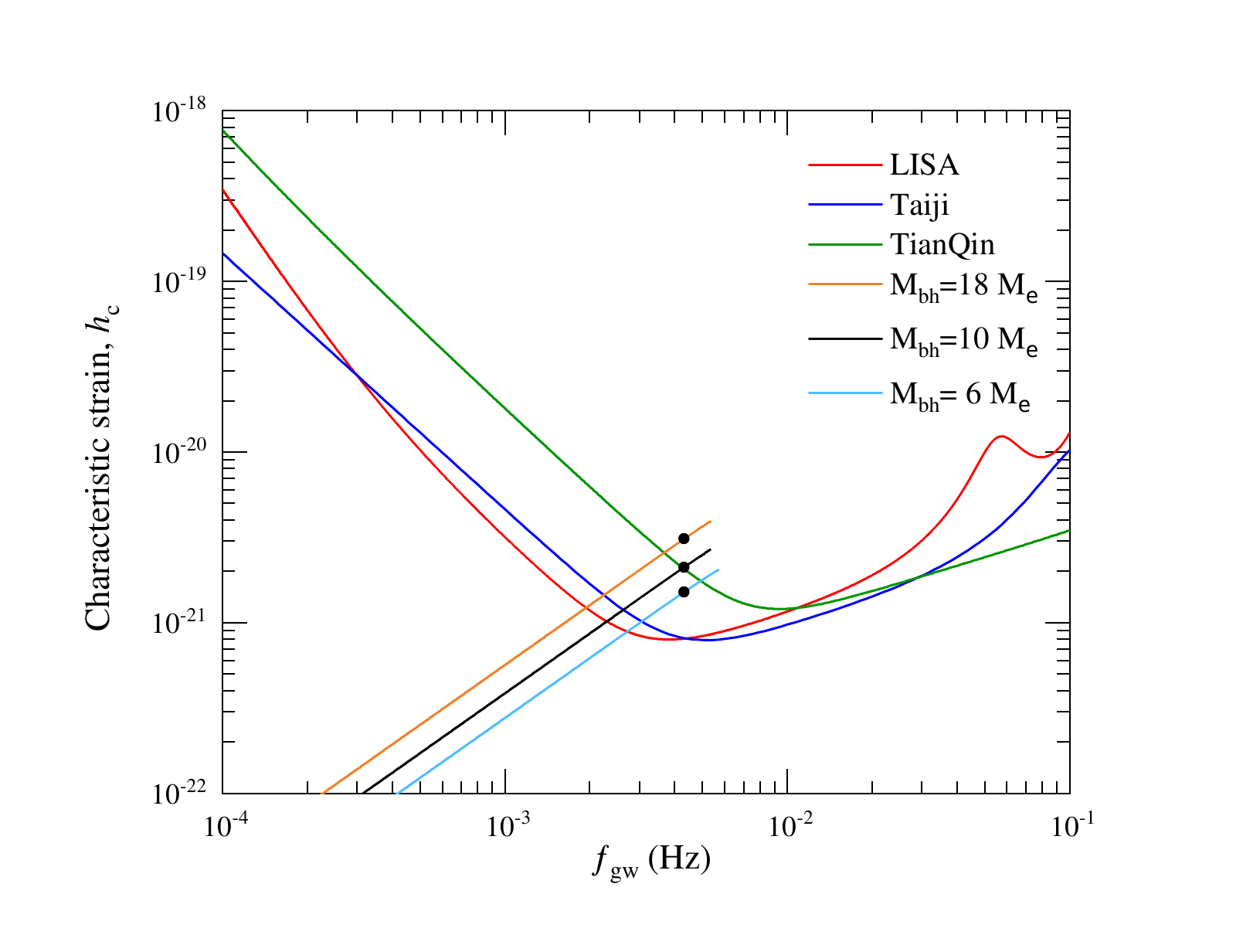}
\caption{Evolution of three BH UCXBs that evolve from BH-MS binaries consisting a BH and a $1.5 ~M_\odot$ MS companion in an initial orbit of $1.0~\rm day$ in the characteristic strain amplitude vs. GW frequency diagram. The red, blue, and green curves represent the sensitivity curves of LISA \citep{amar23}, Taiji \citep{ruan20}, and TianQin \citep{wang19}, respectively. The black, orange, and light blue lines correspond to BHs with initial masses of 10, 18, and $6~M_\odot$, respectively. The solid circles denote the current positions of the BH UCXBs like Seq.1.} \label{LISA}
\end{figure}

In Figure \ref{LISA}, we plot the evolutionary tracks of three BH-MS binaries consisting of a BH (with a mass of $6$, $10$, and $18~M_\odot$) and a $1.5~M_\odot$ donor star in an initial orbit of $1.0~\rm days$ in the characteristic strain amplitude versus GW frequency diagram. The slopes of three tilted lines are $\Delta{{\rm log}h_{\rm c}}/\Delta{{\rm log}(f_{\rm gw}/\rm Hz)}=7/6$, which corresponds to detached BH-WD systems. The characteristic strains of GW signals strongly depend on the initial BH masses. The current GW frequency is $f_{\rm gw}=2/P_{\rm orb}=4.3~mHz$ (see also the solid circles in Figure \ref{LISA}) if Seq.1 is a BH UCXB. LISA and Taiji can detect low-frequency GW signals from these sources even if the BH has a low initial mass of $6~M_\odot$. Due to relatively short arms, it is impossible for TianQin to discover low-frequency GW signals from BH UCXBs if the initial mass of the BH is smaller than $10~M_\odot$. Therefore, the observation of low-frequency GW signals provides an alternative way to confirm whether Seq.1 is a BH UCXB.

\subsection{Constraint for the BH mass}
In table 1, the donor-star masses are distributed in a narrow range of $0.157-0.168~M_\odot$ when the first mass transfer ends. This mass range is related to the narrow range ($0.271-0.436~\rm days$) of orbital periods when donor stars decouple from their Roche lobes. Subsequently, the stripped He cores evolve into He WDs via a contraction and cooling phase. For BH-WD binaries, only those systems with orbital periods shorter than $0.439~\rm days$ can evolve into UCXBs like Seq.1 via a GR within a Hubble time. 

The chirp mass of a detached BH-WD system can be determined by
\begin{equation}
\mathcal{M}=\frac{c^3}{G}\left(\frac{5\dot{f}_{\rm gw}}{96\pi^{8/3}f_{\rm gw}^{11/3}}\right)^{3/5},
\end{equation}
where $\dot{f}_{\rm gw}$ is the GW frequency derivative. If the chirp mass of those detached BH-WD binaries can be accurately measured, the BH masses
will be well constrained from the chirp mass equation (7) due to a narrow range of WD masses. In detached NS-WD binaries as the progenitors of UCXBs, the detailed binary evolution model also found that WD masses are in a narrow range of $0.16-0.17~M_\odot$ \citep{taur18,chen20b}. Therefore, this method can precisely constrain the NS masses within an uncertainty of $\sim 4\%$ for those pre-UCXB systems whose chirp masses can be accurately detected in the future \citep{taur18}.

\subsection{Long gamma-ray bursts without supernova association}
Long gamma-ray bursts (GRBs) are generally thought to be explosive events due to the collapse of massive stars, hence they naturally associated with supernovae \citep{woos93,woos06}. However, some low-redshift long GRBs such as GRBs 060505, 060614, and 111005A were not detected to be associated with supernovae. \cite{dong18} proposed an alternative channel to produce long GRBs without supernova association, which originates from unstable
and extremely violent accretion in a compact BH-WD binary. During the unstable mass transfer, the WD would be disrupted by the BH within several orbital periods \citep{frye99}. According to the WD mass-orbital period relation $P_{\rm orb}=47.2~{\rm s}/(M_{\rm wd}/M_\odot)$ in UCXBs \citep{chen20b}, the orbital periods of our simulated BH-WD systems with a mass transfer are $\sim300~\rm s$. As a consequence, the accretion timescale in such a BH-WD binary with an unstable mass transfer is roughly $\sim10^3~\rm s$.

However, a BH-WD binary with $M_{\rm bh}=10~M_\odot$ and $M_{\rm wd}=0.16-0.17~M_\odot$ should experience a stable mass transfer for an outflow fraction $f=0.9$ in the mass loss and a dimensionless parameter $\lambda=0.99$ \citep[$\lambda$ is the ratio of the specific angular momentum of outflows to $j_{\rm L1}-j_{\rm d}$, in which $j_{\rm L1}$ and $j_{\rm d}$ are the specific angular momentum of the inner Lagrange point and the accretion disk, respectively, see also Figure 4 of][]{dong18}. As a periodic X-ray source, and an absence detecting long GRBs, Seq.1 should be experiencing a stable mass accretion. Our detailed binary evolution model also confirms this point. Therefore, it implies that the BH mass should be $\la10~M_\odot$ according to Figure 4 in \cite{dong18} if Seq.1 is a BH UCXB. 

\subsection{A possible tidal disruption event}
Because the system Seq.1 gradually spirals in due to the angular momentum loss driven by GR, the WD may enter the tidal radius of the BH, which is given by \citep{hopm04}
\begin{equation}
R_{\rm t}=R_{\rm wd}\left(\frac{M_{\rm bh}}{M_{\rm wd}}\right)^{1/3}.
\end{equation}
If so, a tidal disruption event (TDE) will be expected. The radius of a WD can be written as \citep{rapp87,tout97}
\begin{equation}
R_{\rm wd}=0.0115R_{\odot}\sqrt{\left(\frac{M_{\rm ch}}{M_{\rm wd}}\right)^{2/3}-\left(\frac{M_{\rm wd}}{M_{\rm ch}}\right)^{2/3}}, \label{rwd}
\end{equation}
where $M_{\rm ch}=1.44~M_{\odot}$ is the Chandrasekhar mass limit. As $M_{\rm wd}\ll M_{\rm ch}$, the WD radius can be approximately estimated by
\begin{equation}
R_{\rm wd}\approx 0.0115R_{\odot}\left(\frac{M_{\rm ch}}{M_{\rm wd}}\right)^{1/3}. 
\end{equation}

In Seq.1, there still exists a mass transfer, which gives rise to an orbital expansion. Because the GR dominates the orbital evolution of Seq.1, we adopt the GR-driven mass-transfer rate as \citep{king20}
\begin{align}
\dot{M}_{\rm wd}=-1.4\times10^{-5}\left(\frac{M_{\rm bh}}{10~M_\odot}\right)^{2/3}\left(\frac{M_{\rm wd}}{0.16~M_\odot}\right)^2\nonumber\\
\left(\frac{1.5~\rm minutes}{P_{\rm orb}}\right)^{8/3}~M_\odot\,\rm yr^{-1}.   
\end{align}
Ignoring the mass loss, the mass transfer produces an orbital period derivative as
\begin{equation}
\dot{P}_{\rm mt}=\frac{3\dot{M}_{\rm wd}P_{\rm orb}}{M_{\rm wd}}  (\frac{M_{\rm wd}}{M_{\rm bh}}-1).
\end{equation}
However, the GR results in a negative orbital period derivative as
\begin{equation}
\dot{P}_{\rm gr}=-\frac{96G^3}{5c^5}\frac{M_{\rm bh}M_{\rm wd}(M_{\rm bh}+M_{\rm wd})}{a^4}P_{\rm orb},
\end{equation}
where $a$ is the orbital separation. According to Equations (13) and (14), the orbital period derivative during the evolution of Seq.1 prior to a TDE is $\dot{P}=\dot{P}_{\rm gr}+\dot{P}_{\rm mt}$. If $\dot{P}<0$, the orbit of Seq.1 still experiences a rapid decay, and a TDE will be expected in the future.

Assuming that Seq.1 is a BH-WD system with $M_{\rm bh}=10~M_\odot$, $M_{\rm wd}=0.16~M_\odot$, and $P_{\rm orb}=7.7~\rm minutes$, we perform a numerical calculation for the future evolution of Seq.1. If $a<R_{\rm t}$, a TDE is thought to trigger. Our results show that a TDE will occur after 0.12 Myr if Seq.1 is a BH UCXB with $M_{\rm wd}=0.16~M_\odot$ and $M_{\rm bh} = 10~M_\odot$. The main parameters of the system are $M_{\rm bh}=10.05~M_\odot$, $M_{\rm wd}=0.11~M_\odot$, and $P_{\rm orb}\approx2.3~\rm minutes$ when the TDE occurs.

\subsection{Number ratio between BH LMXBs and BH UCXBs}

In Table 1, the average time interval of BH UCXBs is $\tau_{\rm ucxb}\sim0.3~\rm Myr$, but it is as long as $\tau_{\rm lmxb}=t_{\rm deta}-t_{\rm rlof}\sim1~\rm Gyr$ for BH LMXBs. However, BH LMXBs generally appear as soft X-ray transients in observations \citep{king96}. Considering a duty cycle of $d=0.01$, one expects a number ratio between BH LMXBs and BH UCXBs of $N_{\rm lmxb}/N_{\rm ucxb}=d\tau_{\rm lmxb}/\tau_{\rm ucxb}\approx30$ if these UCXBs and LMXBs evolved from BH-MS binaries. In the Galaxy, there exist about 15 BH LMXBs \citep{deng24}, but the unique BH UCXB candidate is the luminous X-ray source X9 in the globular cluster 47 Tucanae \citep{mill15}. Therefore, our simulated time intervals for the BH LMXBs and BH UCXBs are approximately compatible with the observations.

\section{Summary}
In the bulge of M31, the Chandra observations discovered a periodic X-ray source Seq.1 with a period of $7.7~\rm minutes$ and a maximum X-ray luminosity $L_{\rm X}=1.09^{+0.02}_{-0.01}\times10^{38}~\rm erg\,s^{-1}$ in the $0.5-8$ keV band \citep{Zhang24}. However, the detailed stellar evolution model adopting the standard MB law indicate that those BH-MS binaries can only evolve into BH UCXBs with a minimum orbital period of $\sim8.3~\rm minutes$ \citep{qin23}. In this work, we attempt to diagnose the possibility of Seq.1 as a BH UCXB with an orbital period of $7.7~\rm minutes$. Using the MESA code, our detailed stellar evolution models confirm that isolated BH-MS binaries can evolve toward BH UCXBs with an orbital period of $7.7~\rm minutes$ if we adopt the CARB MB law derived by \cite{Van19}. Those BH-MS binaries first evolve into detached BH-WD binaries, then the GR drives a rapid orbital shrinkage and the second mass transfer. In the second short-term episode of the mass transfer, the WDs transfer the material onto the BHs at a rate of $\ge 10^{-9}~M_\odot\,\rm yr^{-1}$ ($\ge 10^{-8}~M_\odot\,\rm yr^{-1}$) in a timescale of $0.16~\rm Myr$ ($0.07~\rm Myr$) when $M_{\rm d,i}=1.25~M_\odot$. If the mass-transfer rate is $1.7\times 10^{-8}~M_\odot\,\rm yr^{-1}$, it would produce an X-ray luminosity of $L_{\rm X}=1.0\times10^{38}~\rm erg\,s^{-1}$, which is compatible with the maximum X-ray luminosity detected in Seq.1.

Those BH-MS binaries with an initial donor-star mass in the range of $1.0-1.8~M_\odot$ and an initial orbital period in the range of $0.6-4.1~\rm days$ could potentially evolve toward BH UCXBs like Seq.1. If the initial periods are not too long, the CARB MB mechanism can still drive those BH-MS binaries with $M_{\rm d,i}=1.6-1.8~M_\odot$ to evolve into Seq.1-like systems once the donor stars develop a convective envelope. To evolve into mass-transferring BH-WD systems, the WD masses concentrate in a narrow range of $0.157-0.168~M_\odot$. As a BH UCXB, the low-frequency GW signals from Seq.1 could be detected by future space-borne GW detectors such as LISA, TianQin, and Taiji. If the chirp mass of this source can be accurately measured, the BH mass will be constrained well from the chirp mass equation because of a narrow range of the WD mass. Furthermore, we expect a tidal disruption event after 0.12 Myr if Seq.1 is a BH UCXB with $M_{\rm wd}=0.16~M_\odot$ and $M_{\rm bh} = 10~M_\odot$.
\\
\\


We thank the referee for a very careful reading and constructive comments that have led to the improvement of the manuscript. This work was partly supported by the National Key Research and Development Program of China (No. 2022YFC2205202), the Major Science and Technology Program of Xinjiang Uygur Autonomous Region (No. 2022A03013-1), the National Natural Science Foundation of China (under grant No. 12273014), and the Natural Science Foundation (under grant number ZR2021MA013) of the Shandong Province.

\bibliography{ref}
\bibliographystyle{aasjournal}
\end{document}